\documentclass[aip,jcp,reprint,groupedaddress]{revtex4-1}
\usepackage[utf8]{inputenc}
\usepackage{amsmath} 
\usepackage{amsfonts}
\usepackage{bm}
\usepackage{amssymb}
\usepackage{graphicx}  
\usepackage{braket}
\usepackage{grffile}
\usepackage{dsfont}
\usepackage{comment}
\usepackage{hyperref}
\usepackage{dcolumn}

\newcommand{\HH}[0]{\mathcal{H}}
\newcommand{\mk}[1]{\overline{#1}}
\newcommand{\nmk}[1]{\breve{#1}}
\newcommand{\KK}[0]{\mathcal{K}}
\newcommand{\LL}[0]{\mathcal{L}}
\newcommand{\I}[0]{\mathcal{I}}
\newcommand{\GG}[0]{\mathcal{G}}
\newcommand{\TC}[0]{_{\rm T}}
\newcommand{\SC}[0]{_{\rm S}}
\newcommand{\PP}[0]{\mathcal{P}}
\newcommand{\QQ}[0]{\mathcal{Q}}
\newcommand{\VV}[0]{\mathcal{V}}
\newcommand{\dd}[2]{\frac{d{#1}}{d{#2}}}
\newcommand{\etadiff}[0]{\eta_\Delta}

\newcommand{\tr}[0]{\text{tr}}
\newcommand{\proj}[2]{| #1 \rangle \! \langle #2 |}
\newcommand{\bbra}[1]{\langle\!\langle{#1}|}
\newcommand{\kket}[1]{|{#1}\rangle\!\rangle}
\newcommand{\bbrakket}[2]{\langle\!\langle{#1}|{#2}\rangle\!\rangle}

\begin{document}

\title{Excitation energy transfer efficiency: equivalence of transient and stationary setting and the absence of non-Markovian effects}

\author{Simon Jesenko}
 \email{simon.jesenko@fmf.uni-lj.si}
\author{Marko \v{Z}nidari\v{c}}%

\affiliation{
Department of Physics, Faculty of Mathematics and Physics, University of Ljubljana
}%

\date{\today}

\begin{abstract}
We analyze efficiency of excitation energy transfer in photosynthetic complexes in transient and stationary setting. In the transient setting the absorption process is modeled as an individual event resulting in a subsequent relaxation dynamics. In the stationary setting the absorption is a continuous stationary process, leading to the nonequilibrium steady state. We show that, as far as the efficiency is concerned, both settings can be considered to be the same, as they result in almost identical efficiency. We also show that non-Markovianity has no effect on the resulting efficiency, i.e., corresponding Markovian dynamics results in identical efficiency. Even more, if one maps dynamics to appropriate classical rate equations, the same efficiency as in quantum case is obtained.
\end{abstract}

\maketitle

\section{Introduction}

Excitation energy transfer in the initial stages of photosynthesis has gained large interest due to coherent beatings observed in two-dimensional electronic spectroscopy experiments on photosynthetic complexes\cite{Lee2007,Engel2007,Collini2010, Panitchayangkoon2010}. In light of these observations, multiple mechanisms have been proposed that could lead to improved efficiency of energy transfer\cite{Rebentrost2009a, Ishizaki2010PCCP, Pachon2011, Ishizaki2009,Plenio2008, Rebentrost2009,Mohseni2008,Caruso2009,Wu2010,Pachon2012}. However, the relevance of the experiments (and suggested mechanisms) for the actual processes \textit{in vivo} is still debated\cite{Brumer2012,Kassal2013,Mancal2010,Cheng2009,Fassioli2012,Tiersch2012,Pachon2013}, as photosynthesis takes place in natural conditions of incoherent continuous sunlight illumination, while the experiments are conducted by a coherent pulsed laser light.

The process of excitation energy transfer (EET) involves electronic excitations on pigments and molecular vibrations of pigments and nearby proteins\cite{may2011charge}. Proper treatment of \textit{vibrational} degrees of freedom (environment) in a description of EET is not trivial, as coupling strengths in pigment-protein complexes (PPCs) are such that the environmental effects can not be treated perturbatively. While in the limit of weak and strong environmental coupling Redfield and F\"orster theory\cite{may2011charge} give simple and intuitive description of excitation dynamics, there are many suggested methods that are trying to properly account for environmental effects also in the intermediate regime\cite{Ishizaki2009a, Huo2010,Ritschel2011,McCutcheon2011,Nalbach2011}. Recently, hierarchical equations of motion\cite{Ishizaki2009a,Tanimura1989,Shi2009} (HEOM) gained much popularity in the context of EET\cite{Ishizaki2009,Zhu2011,Fassioli2012,Dijkstra2012}, as it is formally exact, however, at the expense of high numerical effort\cite{Kreisbeck2011}. Also, due to involved mathematical structure, it offers little insight into underlying principles governing the dynamics of EET.

Two different settings for the EET can be considered. In experiments with short laser pulses the excitation transfer is just a transient phenomenon -- an initial excitation is either transferred to the target site or dissipated in the environment\cite{Cheng2009}. After long time there are no excitations nor currents present. Such a situation will be called a transient setting. In natural conditions though there is a constant flux of incoming photons that continuously create excitations. After a very short transient time a stationary state is established, the so-called nonequilibrium steady state\cite{Brumer2012,Kassal2013}, supporting time-independent energy flow. This second situation will be called a stationary setting.

In this work the focus is on a comparison of a transient and stationary setting, in particular on the differences in the \textit{efficiency} of EET. Efficiency\cite{Ritz2001,Rebentrost2009} corresponds to the probability that the absorption event will result in the energy being transported to the target site. We study two settings because they are physically relevant, i.e., transient case for the pulsed light vs. stationary in the case of natural light. Also, because the stationary setting is by definition time-independent it enables for an easier discussion of the role played by various non-Markovian and oscillatory effects. The difference between the efficiency in the transient and stationary setting is in all relevant situations found to be negligible. Therefore, as far as the efficiency goes, the two settings are equivalent. Not least, it turns out that the stationary setting can also have some advantage in terms of computational speed over the transient setting where the whole time evolution has to be computed.

We consider various approximations when analyzing the efficiency, each providing description at a different level of detail. We start with a generalized quantum master equation, which provides a complete description of EET dynamics, including non-Markovian effects due to the interaction with environment. The kernel for a generalized master equation is obtained from the HEOM method. From the generalized quantum master equation we obtain the corresponding Markovian quantum master equation, and, following the Nakajima-Zwanzig formalism\cite{breuer2002theory}, also the corresponding classical master equation. We shall show that the efficiency is \textit{identical} in all three cases, i.e., for the HEOM, Markovian approximation, as well as for simple classical rate equations. Also, main features of the EET dynamics are retained at each level of approximation. This result suggests that simple rate equations might be adequate for the description of the processes relevant for the biological function of PPCs provided the calculation of rates properly takes into account the underlying quantum mechanics.

\section{Model}

Dynamics of excitations in photosynthetic complexes can be described at different level of detail, and can be either based on derivation from microscopic picture, or phenomenological with parameters obtained from experiments. First we will classify equations of motion (EOMs) based on their mathematical structure, ignoring underlying microscopic model. We will also introduce a formalism that enables a consistent mapping of EOMs from full quantum description to the level of classical rate equations. In the following subsection relevant microscopic model for PPCs is introduced, providing full quantum description of PPCs based on the HEOM method. Note, however, that the finding about the equivalence of efficiencies of EET and the role of non-Markovianity does not depend on the specific form of the microscopic model used.

\subsection{Types of EOMs}

Microscopic description of the photosynthetic system is given by the total density matrix of the system $R(t)$, containing electronic and vibrational degrees of freedom (DOF) of pigments and surrounding proteins. Evolution of $R$ is governed by the Schr\"odinger equation \begin{equation}
\dd{R(t)}{t}=-\frac{i}{\hbar}[\HH,R(t)].
\label{eq:eom_whole_system}
\end{equation}
Such complete description is however computationally intractable due to large number of DOFs. Therefore, the total system is usually divided to a relevant (system) and an irrelevant (environment) part, with the relevant part corresponding to electronic DOFs and the irrelevant to the vibrational DOFs. Then the effective EOMs are derived for the system density matrix $\rho(t)$ only. The procedure is formally exact by Nakajima-Zwanzig formalism\cite{breuer2002theory} by introducing projection operators for the relevant and irrelevant part $\PP$ and $\QQ$ that act on the total density operator, where $\PP$ is chosen such that $\rho(t)\otimes\rho_{\rm ph}=\PP R(t)$. Projectors satisfy usual relations
$\PP^2 = \PP$,
$\QQ^2 = \QQ$ and
$\PP+\QQ = \I$.
When the initial state $R(0)$ and the projector $\QQ$ are such that $\QQ R(0) = 0$, the following equation is obtained,
\begin{equation}
  \dd{\rho(t)}{t}=\int_{0}^{t} \KK(t-\tau) \rho(\tau) d\tau,
  \label{eq:generalized_master_eq}
\end{equation}
which is known as a \textit{generalized quantum master equation}, and contains only the relevant density matrix of electronic DOF. However, calculation of the kernel $\KK(t)$ from the microscopic picture of eq. \eqref{eq:eom_whole_system} is highly nontrivial. Nonetheless, for certain cases of system-environment interaction, efficient numerical schemes have been developed, enabling an exact evolution of system density matrix. Most frequently used in the context of EET are the \textit{hierarchical equations of motion} (HEOM), which are also used in the present paper. In HEOM the direct evaluation of memory kernel $\KK(t)$ and time-nonlocal evolution is circumvented by the introduction of auxiliary operators. The details of the method will be given in next subsection.

In certain regimes the time-nonlocal equation \eqref{eq:generalized_master_eq} can be simplified by the Markovian approximation in which the kernel is taken to be $\KK = \mk\KK \delta(t)$, i.e., there are no memory effects, resulting in a time-local \textit{quantum master equation},
\begin{equation}
  \dd{\rho(t)}{t} = \mk\KK \rho(t).
  \label{eq:master_eq}
\end{equation}

Quantum Markovian eq. (\ref{eq:master_eq}) can also serve as a staring point for the derivation of the corresponding classical dynamics, i.e., equations dictating the evolution of diagonal elements of system's density matrix in a certain basis, $\bm p = (\rho_{00}, \rho_{11}, \ldots, \rho_{nn})$, that is of \textit{populations}. The corresponding \textit{classical generalized master equation} is of the form
\begin{equation}
  \dd{\bm p(t)}{t}=\int_{0}^{t}K(t-\tau) \bm p(\tau) d\tau,
  \label{eq:generalized_master_eq_classical}
\end{equation}
and with an additional Markovian approximation a \textit{classical master equation} is obtained,
\begin{equation}
  \dd{\bm p(t)}{t}= \mk K \bm p(t).
  \label{eq:master_eq_classical}
\end{equation}
Formally, one can derive classical master equation from quantum master equation by employing Nakajima-Zwanzig formalism, where the projection operators $\PP$ and $\QQ$ are chosen to project out only dynamics of populations (see appendix \ref{app:heom_kernel} for details).

In the present work we shall use the term non-Markovian for evolutions governed by a time-dependent kernel, eqs. \eqref{eq:generalized_master_eq} or \eqref{eq:generalized_master_eq_classical}, while we call evolution Markovian if it is determined by a time-local kernel, eqs. \eqref{eq:master_eq} or \eqref{eq:master_eq_classical}.

\subsection{Microscopic model}

Here we specify the microscopic model of PPC that is usually employed when treating EET\cite{may2011charge}. The EOMs derived from the model result in a generalized master equation \eqref{eq:generalized_master_eq}. We start by separating the Hamiltonian into two parts, $\HH=\HH_{\rm ppc} + \HH_{\rm int}$, where $\HH_{\rm ppc}$ corresponds to an isolated PPC (electronic and vibrational DOFs), and $\HH_{\rm int}$ accounts for the electro-magnetic field interaction (leading to absorption/recombination) and interaction with other nearby functional units (e.g. reaction center). In the following, we will treat dynamics due to $\HH_{\rm ppc}$ exactly, while the effect due to $\HH_{\rm int}$ will be treated approximately on a phenomenological level.

Hamiltonian for the isolated PPC is decomposed as
\begin{equation}
  \HH_{\rm ppc} = \HH_{\rm el} + \HH_{\rm ph} + \HH_{\rm el-ph},
  \label{eq:total_hamiltonian}
\end{equation}
with
\begin{align}
  \HH_{\rm el} &= \sum_{m=1}^N \epsilon_m \proj{m}{m} + \sum_{m \neq n = 1}^N V_{mn} \proj{m}{n},\\
  \HH_{\rm ph} &= \sum_{m=1}^N \HH^m_{\rm ph} = \sum_{m=1}^N \sum_\xi \hbar \omega_{m\xi} b_{m\xi }^\dagger b_{m\xi},\\
  \HH_{\rm el-ph} &= \sum_{m=1}^N \HH^m_{\rm el-ph} = \sum_{m=1}^N \sum_\xi g_\xi^m (b_{m\xi}^\dagger + b_{m\xi}) \proj{m}{m},
\end{align}
where $N$ is the number of pigments in PPC, $\HH_{\rm el}$ corresponds to electronic DOFs within single-excitation manifold, $\HH_{\rm ph}$ are phonon DOFs due to pigment and protein vibrations, and $\HH_{\rm el-ph}$ account for exciton-phonon interactions. $\ket{m}$ corresponds to the excitation on the $m$th pigment within the single-excitation subspace, $\epsilon_m$ is the corresponding on-site energy and $V_{mn}$ accounts for the inter-pigment interaction. $b^\dagger_{m \xi}$ and $b_{m \xi}$ are creation/annihilation operators for the $\xi$th phonon mode coupled to the $m$th pigment, $\omega_{m\xi}$ is the frequency of the corresponding mode, and $g_\xi^m$ the coupling of the excitation on the $m$th site to the $\xi$th mode.

Formal solution of eq.~\eqref{eq:eom_whole_system} for the system density matrix in the case of separable initial condition $R(0) = \rho_0 \otimes \rho_{\rm ph} \otimes \rho_{\rm int}$ is given by
\begin{equation}
  \rho(t) = \tr_{\rm ph, int} \big\{ \exp[(\LL_{\rm ppc} + \LL_{\rm int})  t ] \rho_{\rm ph} \otimes \rho_{\rm int} \big\} \rho_0,
\end{equation}
where Liouvillians $\LL$ are linear superoperators determined by their action on a density matrix, $\LL \rho = -\frac{i}{\hbar}[\HH, \rho]$. Evaluation of time evolution of $\rho(t)$ is nontrivial already in the case of an isolated PPC as the pigment-protein interaction cannot be treated perturbatively. Introduction of interaction Hamiltonian $\HH_{\rm int}$ complicates matters even further, as generally $[\HH_{\rm ppc}, \HH_{\rm int}] \neq 0$. For the isolated PPC, exact nonperturbative method has been developed that accounts for the $\HH_{\rm el-ph}$ interaction by introducing a hierarchy of equations of motion (HEOM)\cite{Ishizaki2009a,Tanimura1989,Shi2009} for auxiliary DOFs. The HEOM method can be considered to be an exact description for Lorentzian spectral density and will be used as a starting point for various approximations that we explore. Dynamics due to $\HH_{\rm int}$ will be taken into account approximately by extending resulting HEOMs by effective operators obtained from Born-Markov approximation\cite{Kreisbeck2011,Fassioli2012}.

We assume that each pigment is coupled to an independent phonon bath, where the $m$th bath has a Drude-Lorentz spectral density, $J_m(\omega)\sim \sum_\xi (g_\xi^m/\hbar)^2 \delta(\omega-\omega_{m\xi})$, which is
\begin{equation}
  J_m(\omega) = \frac{2}{\hbar} \lambda_m \frac{\gamma_m \omega}{\omega^2 + \gamma_m^2}.
\end{equation}
Spectral density is characterized by a reorganization energy $\lambda_m$ that specifies strength of the interaction between excitons and phonons, and the bath relaxation time $\gamma_m^{-1}$. In high-temperature limit, $kT>\hbar \gamma$, which is relevant for PPC dynamics at room temperature, HEOMs are of the form\cite{Fassioli2012}
\begin{equation}
\begin{split}
  \label{eq:heom}
  \dd{\rho_{\bm n}(t)}{t}
  = & \left( \LL_{\rm el} - \sum_{j=1}^N n_j \gamma_j \right) \rho_{\bf n} - \sum_{j=1}^N \sum_{k=1}^\infty \frac{c_{jk}}{\nu_{k}}\left[\VV_j,\left[\VV_j,\rho_{\bf n}\right]\right] \\
    & + i \sum_{j=1}^N \sqrt{(n_j + 1)|c_{j0}|} \left[\VV_j,\rho_{\bm n_j^+}\right] \\
    & + i \sum_{j=1}^N \sqrt{\frac{n_j}{|c_{j0}|}}\left(c_{j0} \VV_j \rho_{\bf n_j^-} - c_{j0}^* \rho_{\bf n_j^-} \VV_j \right),
\end{split}
\end{equation}
where $\rho_{\bf n}$ are \textit{auxiliary} density matrices, accounting for memory effects in evolution, and ${\bf n}$ is a vector enumerating them, ${\bf n} = (n_1,n_2,\ldots,n_N)$. System density matrix corresponds to $\rho(t)\equiv \rho_{\bf n = 0}(t)$. Formally, we can represent HEOM as linear first order differential equation $d {\bm \rho(t)}/dt = \mathcal{A} {\bm \rho}(t)$, where $\bm{\rho}$ contains all auxiliary density matrices $\rho_{\bm n}$. We will refer to the sparse operator $\mathcal{A}$ as the HEOM operator. Hierarchy of equations is terminated by a criterion $\sum_i n_i \leq N_{\rm max}$, where $N_{\rm max}$ must be chosen such that the memory effects of the evolution are appropriately accounted for. ${\bf n_j^{\pm}}$ is a shorthand notation for a vector differing from ${\bf n}$ in the $j$th component, $n_j \rightarrow n_j \pm 1$. $\nu_{k} = 2 \pi k / \beta \hbar$ are Matsubara frequencies, and complex coefficients $
c_{jk}$ are given by $c_{j0} = \lambda_j \gamma_j(\cot(\beta \hbar \gamma_j/2) - i)/\hbar$ and $c_{jk} = 4 \lambda_j \gamma_j \nu_k/((\nu_k^2-\gamma_{jk}^2)\beta \hbar^2)$
for $k\geq 1$, where $\beta = 1/(k T)$. We have also introduced a shorthand notation $\VV_j = \proj{j}{j}$. We note that the HEOMs can be represented as a generalized quantum master equation~\eqref{eq:generalized_master_eq} with the procedure for the memory kernel evaluation given in appendix \ref{app:heom_kernel}.

The effect of $\HH_{\rm int}$ can be included into HEOM by introducing an effective time-local Liouvillian $\LL^{\rm eff}$ acting on system density matrix $\rho(t)$. The corresponding combined dynamics can be obtained by augmenting electronic Liouvillian $\LL_{\rm el}$ with the effective interaction Liouvillian, $\LL_{\rm el} \rightarrow  \LL_{\rm el} + \LL^{\rm eff}$, resulting in a hybrid HEOM-Born-Markov set of equations of motion\cite{Kreisbeck2011,Fassioli2012}. The exact form of $\LL^{\rm eff}$ that is used for the modeling of absorption (i.e., pumping), recombination and transfer of excitation to a nearby functional units will be given in the following sections.

\section{Efficiency}
\label{sec:eff}

\begin{figure}
\includegraphics[width=1\columnwidth]{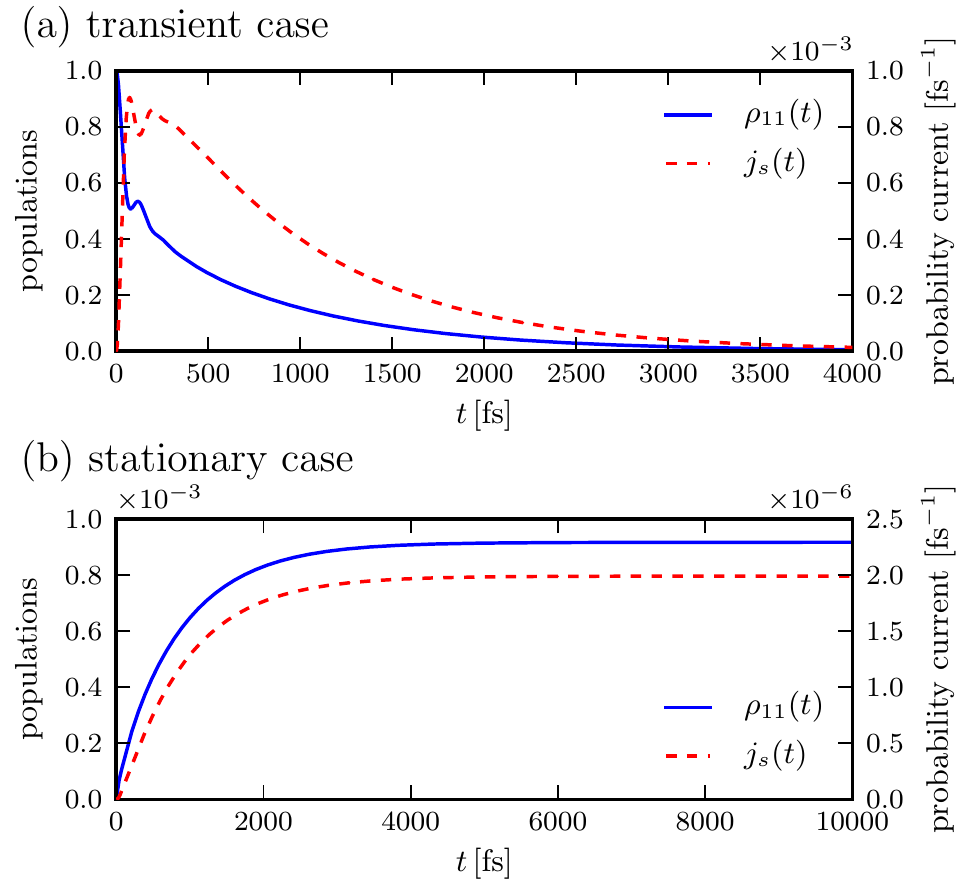}
\caption{\label{fig:plot_trans_stat} An example of time evolution of electronic  density matrix $\rho(t)$ and of probability rate to the sink $j_s(t)$, corresponding to the (a) transient, and (b) stationary setting. Note that in (b) $\rho(t)$ approaches a stationary state, having time-independent $\rho_{11}(t)$ and $j_s(t)$, while in (a) the initial excitation is either transferred to the target site or lost to the environment, resulting in a trivial long-time state. Evolution is shown for a dimer system with parameters $V=100\,\rm{cm^{-1}}$, $\epsilon=100\,\rm{cm^{-1}}$, $\gamma=10^{-2}\,\rm{fs^{-1}}$, $\lambda=100\,\rm{cm^{-1}}$, $\kappa = 1 \times 10^{-3} \,{\rm fs^{-1}}$, $\Gamma = 1 \times 10^{-6}\, {\rm fs^{-1}}$, $\alpha = 1 \times 10^{-6} \,{\rm fs^{-1}}$, $T=300\,\rm{K}$ (see section \ref{sec:dimer}).}
\end{figure}

The efficiency of excitation energy transfer in photosynthesis is the probability that the absorption event will result in a transfer of excitation to the target functional unit, commonly a reaction center. The efficiency of smaller functional unit can also be considered, in which case it corresponds to the probability that the incoming excitation (e.g., due to transfer from an \textit{antennae}) will be transferred to the next functional unit (e.g., a reaction center). Typical example of such smaller functional unit is the Fenna-Mattews-Olson (FMO) complex, which acts as a linker between a chromophoric antennae and a reaction center.

Depending on the setting we use, stationary or transient, the efficiency has to be defined appropriately. In the stationary setting it has to account for the absorption (i.e., pumping) and subsequent transfer of excitation as a continuous stationary process, while in the transient setting one has a time-dependent relaxation dynamics from the initial excited state. Stationary setting is suited for the description of a PPC under natural light conditions, where individual absorption events are not resolved, while the transient one can correspond to the case of absorption due to short light pulse. In previous studies the transient setting has been often employed\cite{Mohseni2008,Plenio2008,Rebentrost2009,Chin2010,Shabani2012,Dijkstra2012,Jesenko2012}. The following analysis demonstrates that the efficiency in the transient and stationary setting is almost identical, with small difference only due to effect of absorption on the internal dynamics of PPC.

While the dynamics of $\rho(t)$ due to phonon bath is exactly treated by previously introduced HEOMs, we still have to specify $\LL^{\rm eff}$ that will be used to model recombination of excitation, transfer to reaction center and in stationary picture also  the absorption (or transfer from an antennae). The relevant system state space consists of single-excitation space $\ket{m}$ and electronic ground state $\ket{0}$. Recombination of the excitation to the ground state will be modeled by the operator
\begin{equation}
  \label{eq:recomb}
  \LL_{\rm{recomb}}(\rho)=2\Gamma \sum_{n=1}^N \left(\proj{0}{n}\,\rho\,\proj{n}{0} - \frac{1}{2}\{\proj{n}{n},\rho \} \right),
\end{equation}
where $\Gamma$ is a site-independent recombination rate. Transfer of excitation to the reaction center is modeled by an analogous operator
\begin{equation}
  \label{eq:sink}
  \LL_{\rm{sink}}(\rho)=2\kappa \left(\proj{0}{s}\,\rho\,\proj{s}{0} - \frac{1}{2}\{\proj{s}{s},\rho \} \right),
\end{equation}
where $s$ denotes a site connected to the reaction center. Note that the reaction center is not explicitly included in $\rho(t)$, and $\LL_{\rm sink}$ only causes transition from sink site to the ground state $\ket{0}$. To model absorption (or transfer from antennae), we introduce an operator\cite{breuer2002theory,Fassioli2012,Manzano2013}
\begin{equation}
  \label{eq:abs}
  \LL_{\rm abs}(\rho) = 2 \alpha \left(\proj{a}{0}\,\rho\,\proj{0}{a} - \frac 1 2 \{\proj{0}{0},\rho \} \right),
\end{equation}
where $\alpha$ denotes the absorption rate, while $a$ is the site that gets excited due to the absorption / transfer process. For simplicity, we have chosen the simplest forms of $\LL_{\rm recomb}$, $\LL_{\rm sink}$ and $\LL_{\rm abs}$, however, they can be trivially generalized to linear combination of operators on different sites, e.g. transfer to sink from multiple sites or absorption on multiple sites.

We shall now define the efficiency in the two settings, stationary and transient. In the \textit{transient setting} the excitation is initialized at a special ``input'' site that we also call the absorption site because it is the same site that is involved in the absorption process in the stationary setting, $\rho_0 = \proj{a}{a}$. The system's state is then propagated with $\LL_{\rm ppc}$ and $\LL\TC^{\rm eff}$, where
\begin{equation}
  \label{eq:L_transient}
  \LL^{\rm eff}\TC = \LL_{\rm recomb} + \LL_{\rm sink}.
\end{equation}
Initial excitation decays to the ground state either due to recombination or due to transfer to the sink. As there is no absorption, the system converges to the ground state $\proj{0}{0}$ after long time. Once time evolution of $\rho(t)$ is obtained, the transient efficiency can be calculated by integrating probability rate of transfer of excitation to the sink,
\begin{equation}
  \label{eq:eff_transient}
  \eta\TC = \int_0^\infty j_s(t) dt = 2 \kappa \int_0^{\infty} \bra{s} \rho(t) \ket{s} dt,
\end{equation}
where the probability rate of transfer to the sink $j_s$ follows from the expression for $\LL_{\rm sink}$. For an example of time evolution in the transient case see Fig.~\ref{fig:plot_trans_stat}a.

For the \textit{stationary setting}, the stationary state of the system $\rho_{\infty} = \rho(t \rightarrow \infty)$ under evolution by $\LL_{\rm ppc}$ and $\LL\SC^{\rm eff}$ is obtained, where
\begin{equation}
  \label{eq:L_stationary}
  \LL\SC^{\rm eff} = \LL_{\rm recomb} + \LL_{\rm sink} + \LL_{\rm abs}.
\end{equation}
Once we have the stationary state $\rho_\infty$ the efficiency is calculated as a ratio between probability rate of transfer to the sink $j_s$ and probability rate of absorption event $j_\alpha$,
\begin{equation}
  \label{eq:eff_stationary}
  \eta\SC = \frac{j_s(\infty)}{j_\alpha(\infty)} =
  \frac{\kappa \bra{s} \rho_\infty \ket{s} }{\alpha \bra 0 \rho_\infty \ket 0 }.
\end{equation}
Time evolution of density matrix populations and probability rates for the stationary scenario is shown in Fig.~\ref{fig:plot_trans_stat}b. Note that the sole difference between the transient (\ref{eq:eff_transient}) and the stationary (\ref{eq:eff_stationary}) setting is in the presence of the $\LL_{\rm abs}$ that causes constant pumping of excitations and the appearance of a nonequilibrium stationary state.

Now we return to the description of dynamics via a generalized master equation~\eqref{eq:generalized_master_eq}, with the kernel $\KK(t)$ corresponding to the microscopic dynamics due to $\LL_{\rm ppc}$ and $\LL^{\rm eff}$. The exact form of the kernel is calculated (see appendix \ref{app:heom_kernel}) using the HEOM-Born-Markov method\cite{Kreisbeck2011,Fassioli2012}. Note that derivations require only specific form of $\LL^{\rm eff}$, while $\LL_{\rm ppc}$ can be arbitrary. For such general scenario, we show that both efficiencies, transient and stationary, depend only on the time-integrated kernels $\mk \KK$, while actual time-dependence of kernels has no effect on the efficiency. Also, the difference between stationary and transient efficiency is very small for the typical parameters of EET in photosynthesis, so both measures of efficiency can be considered equivalent.

Before analyzing each efficiency measure we introduce some common tools that are employed in the analysis. For the comparison of efficiencies Laplace transform is used,
\begin{equation}
  \rho(z) = \int_0^\infty e^{-z t} \rho(t) d t,
\end{equation}
resulting in a Laplace-transformed generalized quantum master equation~\eqref{eq:generalized_master_eq} as
\begin{align}
  \rho(z) = & \Omega(z) \rho(0) \\
  \Omega(z) = &(z - \KK (z))^{-1},
  \label{eq:propagator_laplace}
\end{align}
where $\Omega(z)$ is the Laplace transform of a propagator, and $\KK (z)$ is the Laplace transform of a memory kernel. Laplace-transformed quantities are indicated by their argument $z$. At several occasions we will need a Laplace transform of a propagator resulting from a kernel that is a sum of two terms, $\KK (z)=\KK_1(z)+\KK_2(z)$. Writing $\Omega(z)=1/(z-\KK_1(z)-\KK_2(z))=\Omega_1(z)/(1-\KK_2(z)\Omega_1(z))$, we obtain
\begin{equation}
\Omega(z)=\Omega_1(z)+\Omega_1(z)\KK_2(z)\Omega(z).
\label{eq:sumK}
\end{equation}
In addition, using the final value theorem in a situation with a unique nonequilibrium stationary state, the following useful expression for the stationary state $\kket{\rho_\infty}$ can be obtained,
\begin{equation}
\lim_{z \rightarrow 0} z \Omega(z) \kket{\rho_0} = \kket{\rho_\infty}.
\label{eq:ness}
\end{equation}
For convenience we also introduce a Liouville space notation, i.e., Hilbert-Schmidt space of operators, with $\kket{mn}=\proj{m}{n}$, $\bbra{mn} = (\kket{mn})^\dagger$ and a scalar product $\bbrakket{A}{B} = \tr(A^\dagger B)$. In the analysis we shall decompose $\KK$ and the corresponding propagators to various contributions and observe how each term affects the efficiency of EET. We shall also decompose a time-dependent kernel to the effective Markovian contribution $\mk \KK$ and to the non-Markovian contribution $\nmk \KK$,
\begin{align}
  \KK(t) = \mk \KK \delta(t) + \nmk \KK (t)
  \label{eq:KK_mk_nmk}
\end{align}
where the Markovian contribution corresponds to the integrated kernel, $\mk \KK = \int_0^\infty \KK(t) dt$, while the non-Markovian contribution is $\nmk \KK (t) = \KK(t) - \mk \KK \delta(t)$.

\subsection{Transient efficiency}

In the Laplace picture the transient efficiency is expressed as
\begin{align}
  \eta\TC =&
    2 \kappa \lim_{z \rightarrow 0} \bbra{ss}\Omega\TC(z) \kket{aa},
    \label{eq:trans_eff_z}
\end{align}
which follows from the properties of the Laplace transform, and $\Omega\TC(z)$ is the propagator for the kernel $\KK\TC$ via eq. \eqref{eq:propagator_laplace}. Kernel $\KK\TC$ correspond to the dynamics due to pigment-protein interaction $\LL_{\rm ppc}$ and and effective Liouvillians for the transient case, $\LL^{\rm eff}\TC$ of eq.~\eqref{eq:L_transient}.

We decompose the kernel $\KK\TC$ to the Markovian and non-Markovian contribution $\KK\TC = \mk \KK\TC \delta(t) + \nmk \KK\TC(t)$. Propagators are decomposed accordingly using eq. (\ref{eq:sumK}), resulting in $\Omega\TC(z)=\mk \Omega\TC(z) + \mk \Omega\TC (z) \nmk \KK\TC(z) \Omega\TC(z)$, with the obvious notation $\mk \Omega\TC(z)=1/(z-\mk \KK\TC)$. Inserting this expression into the definition of the transient efficiency \eqref{eq:trans_eff_z}, we obtain two contributions to the efficiency, $\eta\TC = \mk \eta\TC + \nmk \eta\TC$, where
\begin{align}
  \label{eq:efficiency_transient}
  \mk \eta\TC& = 2 \kappa \lim_{z \rightarrow 0} \bbra{ss} \mk \Omega\TC(z) \kket{aa} \\
  \nmk \eta\TC & = 2 \kappa \lim_{z \rightarrow 0} \bbra{ss}\mk \Omega\TC(z) \nmk \KK\TC (z) \Omega\TC(z) \kket{aa}.
\end{align}
The Laplace transform of a non-Markovian kernel vanishes for $z=0$ due to $\int_0^\infty \nmk \KK(t) dt  = 0$, and can thus be approximated for small $z$ as $\nmk \KK\TC(z) \approx \nmk \KK\TC^{(1)}\cdot z + \mathcal{O}(z^2),$ and expression for the non-Markovian contribution as
\begin{align}
  \nmk \eta\TC & = 2 \kappa \lim_{z \rightarrow 0} \bbra{ss}\mk \Omega\TC(z) \nmk \KK\TC^{(1)} z \Omega\TC(z) \kket{aa}
\end{align}
Identifying the limiting expression $\lim_{z \rightarrow 0} z \Omega\TC(z) \kket{aa}$ as the stationary state (\ref{eq:ness}) and noting that in the absence of absorption it is equal to the trivial ground state, $\kket{\rho_\infty} = \kket{00}$, as well as $\nmk \KK\TC^{(1)} \kket{00} = 0$, it follows that the non-Markovian contribution in the transient case vanishes, $\nmk \eta\TC = 0.$ The efficiency in the transient case therefore depends only on the Markovian kernel $\mk \KK\TC$,
\begin{equation}
  \eta\TC = \mk \eta\TC,
\end{equation}
i.e., it does not depend on the non-Markovianity which is all contained in $\nmk \KK\TC(t)$.

\subsection{Stationary efficiency}

For the stationary efficiency the steady state $\kket{\rho_\infty}$ is unique and therefore independent of the initial state $\kket{\rho_0}$. Using eq. (\ref{eq:ness}) the stationary efficiency (\ref{eq:eff_stationary}) can be written as
\begin{align}
  \eta\SC =&
     \lim_{z \rightarrow 0} \frac {\kappa \bbra{ss}\Omega\SC(z)\kket{\rho_0}}
          {\alpha \bbra{00} \Omega\SC(z) \kket{\rho_0}}
\end{align}
Stationary state $\kket{\rho_\infty}$ is also the zero-eigenvector of the corresponding Markovian kernel, which is evident by observing the stationarity condition for the generalized master equation \eqref{eq:generalized_master_eq},
\begin{equation}
\frac{d}{d t} \kket{\rho_\infty}=\int_0^\infty \KK(t) \kket{\rho_{\infty}} dt = \mk \KK \kket{\rho_{\infty}} = 0.
\end{equation}
Therefore, using eq. \eqref{eq:eff_stationary}, we can equivalently write the efficiency with the Markovian propagator only,
\begin{align}
  \eta\SC =&
     \lim_{z \rightarrow 0} \frac {\kappa \bbra{ss}\mk \Omega\SC(z)\kket{\rho_0}}
          {\alpha \bbra{00} \mk \Omega\SC(z) \kket{\rho_0}}.
  \label{eq:stat_eff_laplace_nonmarkovian}
\end{align}
Similarly as in the transient case, the efficiency does not depend on the non-Markovian part $\nmk \KK\SC(t)$. To obtain the stationary efficiency we therefore only need $\mk \KK\SC$ and not the full kernel $\KK\SC(t)=\mk\KK\SC \delta(t)+\nmk\KK\SC(t)$. We are going to write $\mk\KK\SC$ as a sum of a transient Markovian kernel $\mk\KK\TC$, the absorption Liouvillian $\LL_{\rm abs}$, and the rest,
\begin{equation}
\mk\KK\SC=\mk\KK\TC+\LL_{\rm abs}+\mk \KK_{\rm abs}^{\rm ph},
\label{eq:stat_kernel_decomposition}
\end{equation}
where (as we shall see small) term $\mk \KK_{\rm abs}^{\rm ph}$ arises due to the non-commutativity $[ \LL_{\rm abs}, \LL_{\rm el-ph}] \neq 0$. Expression for the $\eta\SC$ can now be further simplified using eq.~(\ref{eq:sumK}), by writing the Markovian propagator for the stationary case as a sum of propagators for the transient case and the rest, $\mk \Omega\SC(z) = \mk \Omega\TC(z) + \mk \Omega\TC(z)(\LL_{\rm abs} + \mk \KK_{\rm abs}^{\rm ph}) \mk \Omega\SC(z)$, where we also used the fact that the Laplace transform of a Markovian kernel, being a delta function in time, is equal to the kernel itself. Inserting this expression into the numerator of eq. \eqref{eq:stat_eff_laplace_nonmarkovian}, we obtain
\begin{equation}
    \eta\SC =
     \lim_{z \rightarrow 0} \frac{\kappa}{\alpha} \frac {z \bbra{ss}\mk \Omega\TC(z)(\LL_{\rm abs} + \mk \KK_{\rm abs}^{\rm ph}) \mk \Omega\SC(z) \kket{\rho_0} }
          {z \bbra{00} \mk \Omega\SC(z) \kket{\rho_0}},
    \label{eq:eta_s_intermediate}
\end{equation}
where we have taken into account that the stationary state for the transient setting is trivial, i.e., only ground state is occupied, $\lim_{z \rightarrow 0} \bbra{ss} z \mk \Omega\TC(z) \kket{\rho_0} = \bbrakket{ss}{00} = 0$. After inserting the $\LL_{\rm abs}$ from eq. \eqref{eq:abs}, the expression for the stationary efficiency becomes a sum of the transient efficiency and a correction due to the absorption,
\begin{align}
  \eta\SC = \eta\TC + \etadiff.
  \label{eq:eta_abs}
\end{align}
The correction due to the absorption can be expressed as
\begin{equation}
  \etadiff = 2 \kappa \lim_{z \rightarrow 0}\sum_{m,n=1}^N \bbra{ss}\mk \Omega\TC(z) \mk \KK_{\rm abs}^{\rm ph} \kket{mn}\bbra{mn}\mk \Omega\SC^{\rm ph}(z) \kket{aa},
\end{equation}
with the propagator $\mk \Omega\SC^{\rm ph}(z) = (z - \mk \KK\TC - \mk \KK^{\rm ph}_{\rm abs})^{-1}$.
The details of the calculation can be found in appendix \ref{app:stationary_to_transient}.

Numerical calculations in the following sections show that the difference between the stationary efficiency and the transient efficiency $\etadiff$ is very small for the microscopic model of PPC considered. Therefore, for practical applications, they can be considered to be the same. Observe that if one starts with a Markovian description of PPC dynamics, eq.~(\ref{eq:master_eq}), adding absorption term $\mathcal{L}_{\rm abs}$ to obtain a stationary setting, the efficiency difference $\etadiff$ is exactly zero.

In the analysis above, we have considered transient and stationary efficiency in the case of dynamics described by a generalized master equation \eqref{eq:generalized_master_eq}. As the efficiency only depends on the corresponding Markovian kernel $\mk \KK$, the dynamics under time-local quantum master equation \eqref{eq:master_eq} results in the identical efficiency of EET, $\eta\SC=\mk\eta\SC$ and $\eta\TC=\mk\eta\TC$. That is, a detailed time-dependence, e.g., non-Markovian oscillations in density matrix elements, has no direct effect on the efficiency.

For classical master equation \eqref{eq:generalized_master_eq_classical} and \eqref{eq:master_eq_classical} an analogous analysis can be conducted, where the effective rates must be calculated from the corresponding effective Liouvillians for the transient or stationary case of eq. \eqref{eq:L_transient} or \eqref{eq:L_stationary}. Thus, similarly as in quantum case, for classical EOMs the efficiency also depends only on the classical Markovian kernel $\mk K$. Moreover, if one projects a quantum master equation to a generalized classical master equation in an appropriate basis, and so that the effective Liouvillians translate to the corresponding effective rates, the efficiency of EET is the same in both cases. Therefore, mapping of dynamics from non-Markovian quantum description to Markovian classical description, $\KK(t) \rightarrow \mk \KK \rightarrow K(t) \rightarrow \mk K$, does not affect the efficiency, i.e., all four efficiencies are exactly the same. Mapping of time-independent Markovian quantum master equation with kernel $\mk \KK$ to the corresponding time-dependent non-Markovian classical master equation with kernel $K(t)$ is done by employing the Nakajima-Zwanzig formalism, where a projection operator is chosen such that the diagonal elements of system density matrix represent the relevant subsystem. See appendix \ref{app:heom_kernel} for details.

The above findings about the equivalence of the two efficiency measures and on the absence of non-Markovian effects do not rely on the specific form of the microscopic model, e.g., on the exact form of the spectral density $J(\omega)$. The value of the efficiency itself of course does depend on microscopic parameters\cite{Mohseni2008,Plenio2008,Rebentrost2009,Chin2010} like the spectral density\cite{Nalbach2011,Kreisbeck2012,Kolli2012,Rey2013}, however, $\eta\SC$ and $\eta\TC$ are affected in exactly the same way.

In the following, we shall calculate the transient and stationary efficiencies and the corresponding populations dynamics numerically using the HEOM formalism for the specific microscopic model described in previous section. Equivalence of efficiency measures will be demonstrated as well as mapping of a generalized master equation to a corresponding Markovian classical master equation using the Nakajima-Zwanzig formalism.

\section{Dimer system}
\label{sec:dimer}
\begin{figure}
\includegraphics[width=1\columnwidth]{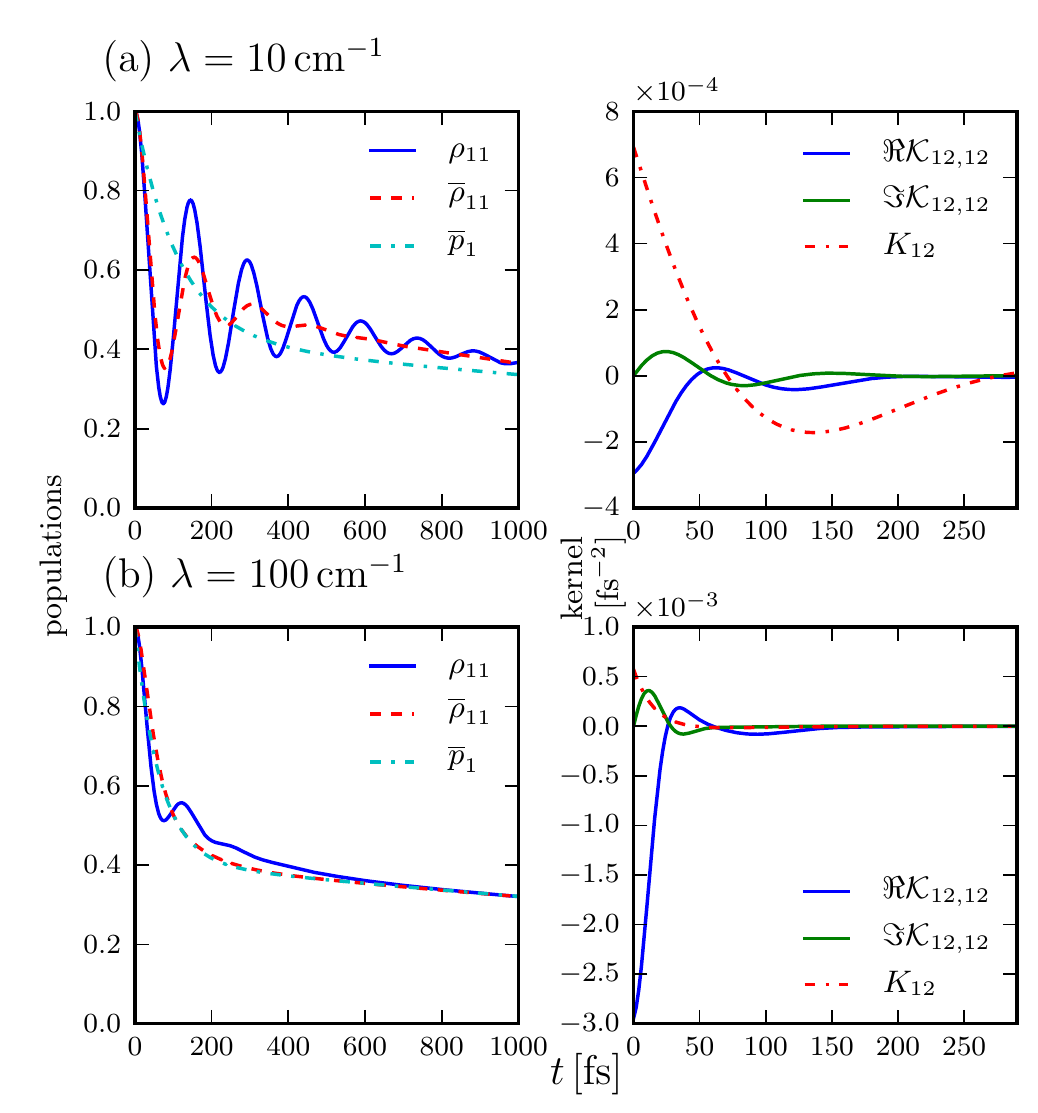}
\caption{\label{fig:dimer_populations_and_kernel}
Time evolution for the transient setting in a dimer system and different levels of approximation: exact HEOM method ($\rho_{11}(t)$, solid line; full kernel $\KK\TC(t)$), Markovian approximation ($\mk \rho_{11}(t)$, dashed line; $\mk \KK\TC$) and classical Markovian rate equations ($\mk p_1(t)$, dot-dashed line; $\mk K\TC$). Evolution of the population of the input site $\rho_{11}(t)$ (initial state is $\rho_0=\proj{1}{1}$) and of the matrix element of a non-singular part of the memory kernel $\KK_{12,12}(t)$ (real and imaginary part) as well as of the classical non-Markovian kernel $K_{12}(t)$ is shown. Two different reorganization energies are used, in (a) $\lambda=10\,{\rm cm^{-1}}$ and in (b) $\lambda=100\,{\rm cm^{-1}}$. Other parameters are $V=100\,\rm{cm^{-1}}$, $\epsilon=100\,\rm{cm^{-1}}$, $\gamma=10^{-2}\,\rm{fs^{-1}}$, $T=300\,\rm{K}$. Note that even thou time dependence is different, the efficiency $\eta\TC$ is the same in all three cases.}
\end{figure}

\begin{figure}
\includegraphics[width=1\columnwidth]{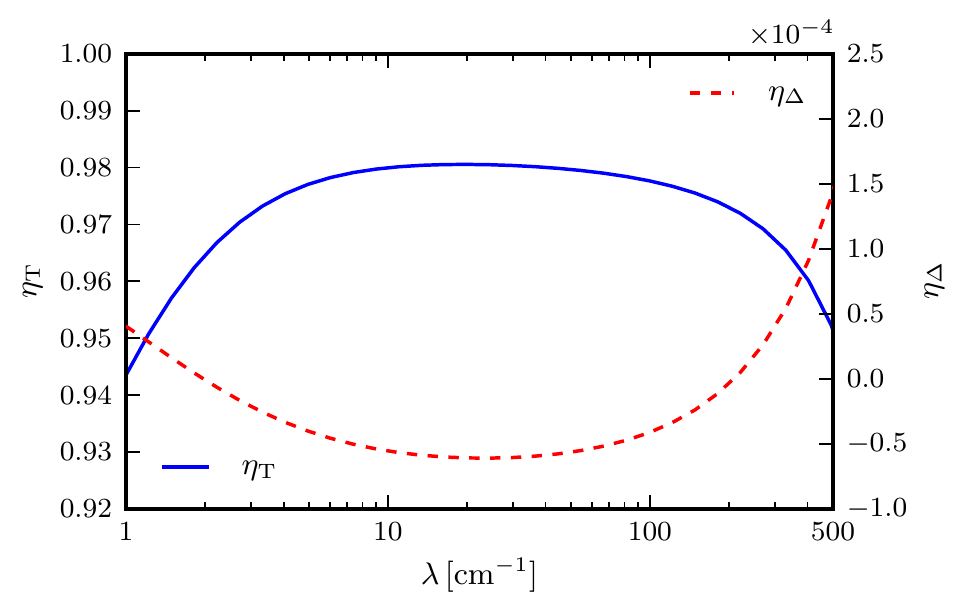}
\caption{\label{fig:plot_dimer_efficiency_vs_lambda} Calculated efficiency of excitation transfer for the dimer system at different values of reorganization energy $\lambda$. Transient efficiency $\eta\TC$ is shown (solid line, left axis) and the absorption contribution in the stationary case $\etadiff$ (dashed line, right axis). Observe that the difference $\etadiff$ is of order $\sim 10^{-4}$ and therefore, for practical purposes, $\eta\TC \approx \eta\SC$. Parameters are the same as in Fig.~\ref{fig:dimer_populations_and_kernel}.}
\end{figure}

As an illustrative example we analyze the case of a two-site PPC. The interaction with environmental phonons is treated exactly within the HEOM formalism from which the time-dependent kernel $\KK(t)$ is explicitly evaluated (see appendix \ref{app:heom_kernel} for more details). Transient efficiencies $\eta\TC$ and stationary efficiencies $\eta\SC$ are calculated for a range of reorganization energies $\lambda$, demonstrating that the difference between the efficiency measures $\etadiff$ is negligible. For the transient case, mapping of dynamics from non-Markovian generalized master equation \eqref{eq:generalized_master_eq} to the corresponding classical master equation \eqref{eq:master_eq_classical} via Nakajima-Zwanzig formalism is also demonstrated.

The total Hamiltonian for the dimer is of the form \eqref{eq:total_hamiltonian} for two sites, $N=2$. The relevant parameters of the electronic Hamiltonian $\HH_{\rm el}$ are the on-site energy difference $\epsilon = \epsilon_{2} - \epsilon_{1}$ and the inter-site interaction strength $V = V_{12}$. Each site is coupled to an independent bath with identical parameters $\lambda_n = \lambda$ and $\gamma_n = \gamma$. First site is an input, $a=1$, while the second is connected to the reaction center, $s=2$. We consider parameters within ranges typical for PPCs. Thus, if not stated otherwise, the temperature is $T=300\,{\rm K}$, bath relaxation time $\gamma=10^{-2}\,\rm{fs^{-1}}$, while recombination rate $\Gamma = 2 \times 10^{-6}\,\rm{fs}^{-1}$ and sink rate $\kappa = 2\times10^{-4}\,\rm{fs}^{-1}$. For the stationary case the absorption rate is taken the same as the relaxation rate, $\alpha = \Gamma$, however, the efficiency $\eta\SC$ (and the corresponding difference from the transient case $\etadiff$) does not depend on the actual choice of $\alpha$.

In Fig.~\ref{fig:dimer_populations_and_kernel}, exact time evolution of the input site population $\rho_{11}(t)$ (starting from the initial state $\rho_0=\proj{1}{1}$) is shown for two values of reorganization energy $\lambda$ in the transient setting. Emergence of incoherent dynamics is evident as the reorganization energy is increased, resulting in a faster decay of coherent oscillations seen at short times. Time dependence of matrix element $\KK_{12,12}(t)$ of a non-singular part of the kernel \eqref{eq:KKns} is also shown, where we use a short notation $\KK_{12,12}(t) \equiv [\KK_{ns}(t)]_{12,12}$. Note that other non-zero matrix elements of $\KK(t)$ also decay on a comparable time scale. Markovian dynamics for the integrated memory kernel $\mk \KK$ is also calculated. Comparing the time dependence of population on the site 1 for the exact non-Markovian evolution and the Markovian approximation we can see that the Markovian evolution results in a faster decay of coherent oscillations. In addition, oscillations for the non-Markovian and Markovian case, although different, are such that the efficiency is the same in all three cases.

We also mapped dynamics from quantum Markovian master equation with kernel $\mk \KK$ to the corresponding classical generalized master equation using Nakajima-Zwanzig formalism, obtaining a time-dependent classical kernel $K(t)$. Time dependent rate $K_{21}(t)$ is shown in Fig.~\ref{fig:dimer_populations_and_kernel}. The dynamics of classical populations $\bm p(t)$, eq.~\eqref{eq:generalized_master_eq_classical}, is identical as the dynamics of the diagonal elements of $\mk \rho(t)$ for quantum Markovian case because the mapping is exact\footnote{Mapping of quantum master equation with kernel $\mk \KK$ to classical generalized master equation with kernel $K(t)$ via Nakajima-Zwanzig formalism is exact only when the initial state $\rho_0$ is diagonal, i.e., no coherences are present. Otherwise, inhomogeneous terms have to be included in classical master equation}. Integrating time-dependent kernel $K(t)$, classical time-independent master equation with kernel $\mk K$ and population dynamics  $\mk{\bm p}(t)$ is obtained, eq.~\eqref{eq:master_eq_classical}. Again, the decay of initial oscillatory dynamics is evident when approximating dynamics of $\mk \rho(t)$ with the corresponding classical Markovian dynamics $\mk{\bm p}(t)$. Here we emphasize that while the populations dynamics under mapping $\mk \rho(t) \rightarrow \mk{\bm p}(t)$ are all different, the resulting efficiency of EET is identical.

For the stationary case (\ref{eq:L_stationary}) exact steady state is obtained by finding the zero-eigenvector of a sparse HEOM operator $\mathcal{A}$ from which the corresponding stationary efficiency $\eta\SC$ is calculated. We have calculated $\etadiff$ for a range of reorganization energies $\lambda$. The results are shown in Fig.~\ref{fig:plot_dimer_efficiency_vs_lambda}. For the parameters taken, the contribution due to the absorption is of order of $\sim 10^{-4}$, indicating that for the purpose of analysis of PPC efficiency, both settings can be considered to be equivalent.

\section{Fenna-Mattews-Olson complex}

\begin{figure}
\includegraphics[width=1\columnwidth]{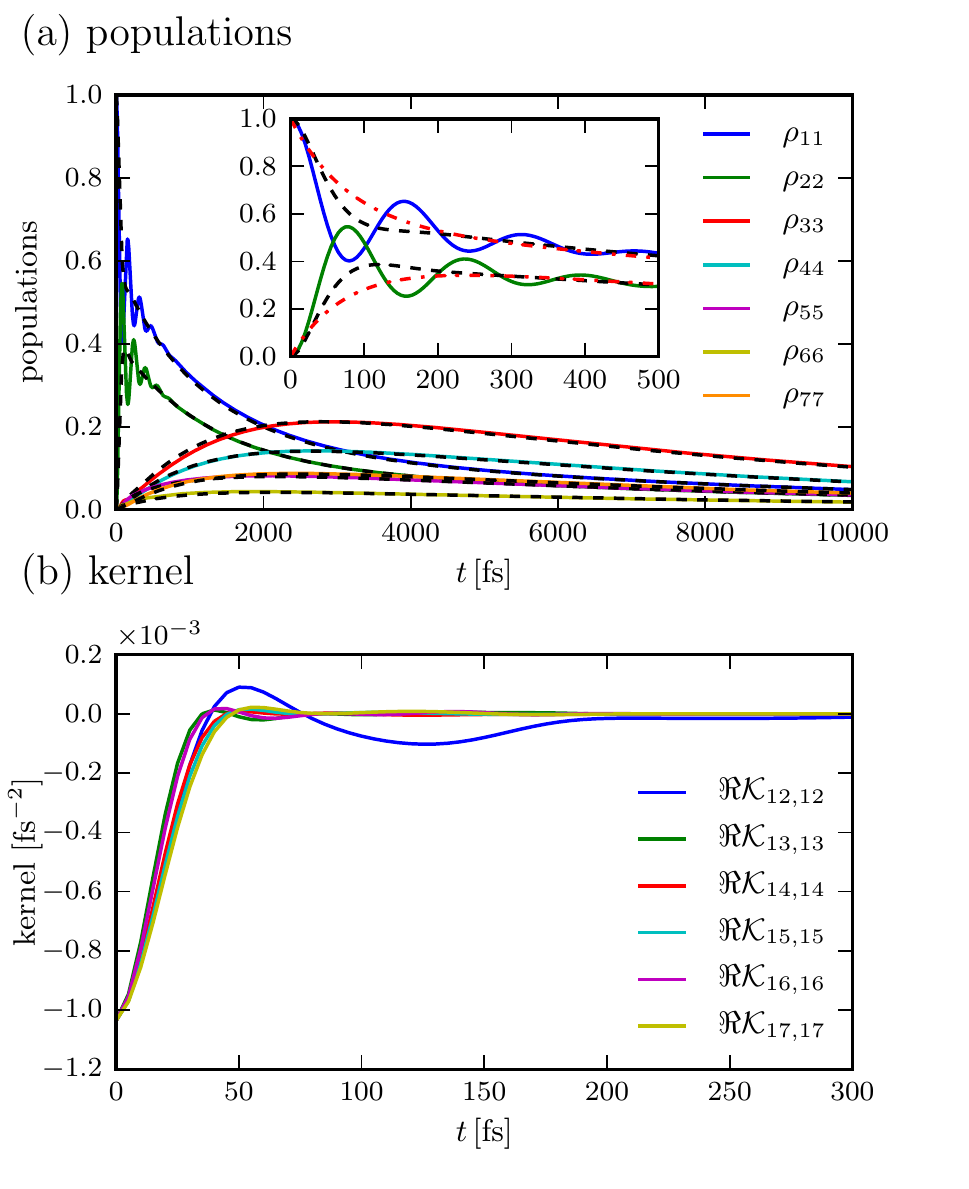}
\caption{\label{fig:plot_fmo_pop_and_kernel} \label{fig:plot_fmo_pop_and_kernel:a} (a) Time evolution of populations $\rho_{mm}(t)$ for the transient evolution of the FMO model and the initial state $\rho_0=\proj{1}{1}$. Dashed lines represent corresponding Markovian evolution $\mk \rho(t)$. In the inset, a short-time dynamics for $\rho_{11}$, $\rho_{22}$, Markovian dynamics $\mk \rho_{11}$ and $\mk \rho_{22}$ (dashed black line) and classical Markovian dynamics $\mk{p}_1$, $\mk{p}_2$ (dash-dotted red line) is shown. (b) Real part of selected elements of a non-singular part of the time-dependent kernel $\KK(t)$ for the FMO model. Parameters of simulation are $\gamma^{-1}=166\,\rm{fs}$, $\lambda=35\,\rm{cm^{-1}}$, $\kappa = 2 \times 10^{-4} \,{\rm fs^{-1}}$, $\Gamma = 2 \times 10^{-6}\, {\rm fs^{-1}}$, $T=300\,\rm{K}$. }
\end{figure}

The Fenna-Matthews-Olson complex (FMO) is often considered in the studies of PPC as its structure is well known and a large number of various experimental studies has been conducted on it\cite{Engel2007,Panitchayangkoon2010,Adolphs2006}. We have used the HEOM method to calculate the transient and stationary efficiencies and to evaluate the memory kernel.

We considered a 7-site FMO model, with electronic Hamiltonian $\HH_{\rm el}$ as specified in Ref. \onlinecite{Adolphs2006}. Site 1 is considered as an input site, while the site 3 is connected to the reaction center. Each site is interacting with an identical phonon bath with parameters $\lambda=35\,\rm{cm^{-1}}$, $\gamma^{-1} = 166\,\rm{fs}$, already used in previous studies \cite{Kreisbeck2011,Ishizaki2009,Brixner2005}. Recombination and sink rates were taken from Ref. \onlinecite{Kreisbeck2011}, with $\Gamma = 2 \times 10^{-6}\, {\rm fs^{-1}}$ and $\kappa = 2 \times 10^{-4}\, {\rm fs^{-1}}$.

Time evolution of populations $\rho(t)$ for the transient case is shown in Fig.~\ref{fig:plot_fmo_pop_and_kernel}a; dashed lines in addition show Markovian dynamics $\mk \rho(t)$ due to $\mk \KK$. For Markovian dynamics the initial oscillatory behavior of populations is not as pronounced as in a dimer, while at latter times Markovian and non-Markovian evolutions result in almost the same site populations. In the inset a short time dynamics for sites 1 and 2 is shown. Additionally, populations $\mk{\bm p}(t)$ for classical master equation, resulting from the mapping $\mk \KK \rightarrow K(t) \rightarrow \mk K$, are also shown (dash-dotted line). At latter times populations $\mk{\bm p}(t)$ approach the values corresponding to the full quantum master equation $\rho(t)$. Note again that the efficiency does not depend on the level of approximation.

To provide some insight on the time-dependent kernel that is relevant in a PPC, few entries of the non-singular part of kernel $\KK_{ns}(t)$ are shown in Fig.~\ref{fig:plot_fmo_pop_and_kernel}b. Plotted entries contribute to the decay of coherences between site 1 and other sites. Note that the element $\KK_{12,12}$ has a larger decay time than other elements, which can be related to the fact that the site 1 is most strongly coupled to the site 2 in $\HH_{\rm el}$.

For the used parameters the difference between the stationary and the transient efficiency is less than the estimated error of the calculation due to the truncation of the HEOM hierarchy at $N_{\rm max} = 8$, with the efficiencies estimated to be $\eta\SC \approx \eta\TC \approx 0.962$, and the truncation error being of the order $\sim 10^{-4}$. We have estimated the truncation error by observing the convergence of stationary efficiency for calculations with HEOM truncation at level up-to $N_{\rm max}=12$. For practical purposes the efficiencies $\eta\TC$ and $\eta\SC$ can thus be considered to be the same.

We note that due to the equivalence of efficiency measures in stationary and transient setting one might choose to calculate the one that is easier to evaluate. Stationary efficiency turns out to be more convenient in that respect, i.e., its calculation can be faster than calculating whole time evolution of $\rho_{\bm n}(t)$, as it requires only the calculation of a stationary state $\rho_\infty$, being the eigenvector with the zero eigenvalue of the HEOM operator $\mathcal{A}$. Taking advantage of the sparsity of $\mathcal{A}$ and using iterative eigenvalue algorithms we obtain the stationary state.

\section{Conclusion}
We have studied two physically relevant settings of EET, namely, a transient situation in which the initial excitations decay to zero, and a stationary setting in which constant pumping causes the system to converge to a nonequilibrium stationary state. Comparing efficiency between the transient and stationary setting we observed that the difference is very small for the exact description, while it is exactly zero for Markovian equations, like, e.g., the Lindblad equation. Equivalence of efficiency measures in transient and stationary case validates findings about efficiencies in PPCs based on observing transient dynamics from specific initial states\cite{Mohseni2008,Plenio2008,Rebentrost2009,Chin2010,Shabani2012,Dijkstra2012,Jesenko2012}. Therefore, the mechanisms leading to higher efficiencies established in the transient picture, such as environment-assisted quantum transport\cite{Mohseni2008,Plenio2008,Rebentrost2009,Chin2010,Nalbach2011,Kreisbeck2012,Kolli2012,Rey2013} and supertransfer \cite{Lloyd2010,Kassal2013}, are also relevant in the case of incoherent light illumination, complying with the findings of Ref.~\onlinecite{Kassal2013}.

In both settings, transient and stationary, the efficiency does not depend on a (time-dependent) non-Markovian part of the kernel. Same result also holds for the description with the classical rate equations. Additionally, if one obtains classical rate equations from full quantum description by employing Nakajima-Zwanzig formalism, the resulting efficiencies are the same as in full quantum description. The only feature of the dynamics that is not reproduced by the classical or quantum Markovian equations is the initial oscillatory motion of populations being present for pure initial states. While physical relevance of evolutions from pure initial quantum states for the \textit{in vivo} process of photosynthesis is questionable\cite{Kassal2013,Brumer2012}, this oscillatory motion, being present or not, does in no way affect the corresponding efficiency of EET, i.e. an approximate dynamics without oscillatory character results in identical efficiencies.

This suggests that the rate equations are adequate for the description of EET in biological processes, as long as the mapping from full non-Markovian quantum picture to the corresponding classical picture is properly treated. This is consistent with previous essentially classical descriptions of EET\cite{Zimanyi2010,Briggs2011}. The procedure for obtaining classical rates from a microscopic model is however highly non-trivial\cite{Cao2009,Wu2012} as the coupling to environmental DOFs can not be treated perturbatively. Starting from the exact quantum description, we use Nakajima-Zwanzig formalism and the Markovian approximation to obtain a description on a relevant subspace. Thus, we obtain, as far as the efficiency is concerned, equivalent descriptions of dynamics at different levels of detail. This also provides a straightforward way of comparing different approximations.

\appendix

\section{Evaluation of memory kernel from HEOM}
\label{app:heom_kernel}

We shall evaluate the memory kernel for system density matrix based on the hierarchical equations of motion. We are treating HEOM as evolution $d \bm{\rho}(t)/dt = \mathcal{A} \bm{\rho}(t)$, where $\mathcal{A}$ is a time-independent linear sparse operator defined by eq. \eqref{eq:heom}. Memory kernel is obtained from $\mathcal{A}$ by projecting out auxiliary degrees of freedom $\rho_{\bm n\neq0}$ using Nakajima-Zwanzig formalism\cite{breuer2002theory}, resulting in a sum of singular and non-singular contribution
\begin{equation}
  \KK(t) = \KK_s \delta(t) + \KK_{ns}(t),
\end{equation}
which can be evaluated in Schr\"odinger picture as
\begin{align}
  \label{eq:KKs}
  \KK_s = & \PP \mathcal{A} \PP,\\
  \label{eq:KKns}
  \KK_{ns}(t) = & \PP \mathcal{A} \GG(t) \QQ \mathcal{A} \PP.
\end{align}
$\PP$ and $\QQ$ are projectors to $\rho_{\bm n=0}$ and $\rho_{\bm n\neq0}$, respectively, and $\GG(t)$ is the propagator for the irrelevant part $\QQ \bm \rho$, \begin{equation}
  \GG(t)=\exp(\QQ\mathcal{A} t),
  \label{eq:propagator_A}
\end{equation}
being a solution of a differential equation,
\begin{equation}
  \frac{d \GG(t)}{dt} = \QQ \mathcal{A} \GG(t),
  \label{eq:propagator_A_diff}
\end{equation}
with the initial condition
\begin{equation}
  \GG(0) = I.
\end{equation}
Numerically, one can evaluate each column of $\GG(t)$ individually as $ d [\GG(t)]_j/dt = \QQ \mathcal{A} [\GG(t)]_j$, where $[\GG]_j$ is the $j$th column of $\GG$. However, direct evaluation of eq. \eqref{eq:propagator_A_diff} becomes intractable when the number of sites $N$ and the hierarchy truncation level $N_{\rm max}$ is increased, even if sparsity of operator $\mathcal{A}$ is taken into account. The number of auxiliary matrices in HEOM is given by\cite{Ishizaki2009} $N_{\rm tot} = (N + N_{\rm max})!/(N!N_{\rm max}!)$, while each auxiliary matrix has $N^2$ elements. Dimensionality of HEOM operator $\mathcal{A}$ is thus $N_{\mathcal{A}}=N^2 N_{\rm tot}$. Obtaining $\GG(t)$ requires numerical solution of $N_{\mathcal{A}}$ differential equations for vectors with $N_\mathcal{A}$ components.

A complete solution of $\GG(t)$ is however not required to obtain the kernel. This is evident if we introduce the operator \begin{equation}
  \mathcal{R}(t) = \mathcal{A}\GG(t) \QQ \mathcal{A} \PP,
\end{equation}
which is calculated by integrating
\begin{equation}
  \frac{d \mathcal{R}(t)}{dt} = \mathcal{A} \QQ \mathcal{A} \GG(t) \QQ \mathcal{A} \PP = \mathcal{A} \QQ \mathcal{R}(t),
\end{equation}
with the initial condition \begin{equation}
  \mathcal{R}(0) = \mathcal{A} \QQ \mathcal{A} \mathcal{P}.
\end{equation}
From the operator $\mathcal{R}$, a non-singular part of the kernel is obtained as \begin{equation}
  \KK_{ns}(t) = \PP \mathcal{R}(t).
  \label{eq:pprr}
\end{equation}
The main advantage of calculating $\mathcal{R}(t)$ instead of $\mathcal{G}(t)$ is that due to a projection in eq.~\eqref{eq:pprr}, only columns of $\mathcal{R}(t)$ that correspond to relevant degrees of freedom have to be calculated when obtaining the kernel $\KK_{ns}(t)$, i.e., only solution of $N^2$ differential equations is required instead of $N^2 N_{\rm tot}$.

When obtaining classical generalized master equation~\eqref{eq:generalized_master_eq_classical} from quantum master equation~\eqref{eq:master_eq}, mapping of a time-independent kernel $\mk \KK$ to a time-dependent kernel $K(t)$ is also obtained by the procedure introduced above. Kernel $\mk \KK$ assumes the role of $\mathcal{A}$, and operators $\PP$ and $\QQ$ are projectors on the diagonal and off-diagonal elements of density matrix, respectively.

\section{Contributions to stationary efficiency}
\label{app:stationary_to_transient}

In this appendix, we start with the expression for the stationary efficiency in Laplace picture, eq. \eqref{eq:eta_s_intermediate}, and show that it can be written as a sum of transient efficiency $\eta\TC$ and contribution due to the effects of phonon environment on the absorption. We start be rewriting absorption Liouvillian from eq. \eqref{eq:abs} as
\begin{equation}
\begin{split}
\LL_{\rm abs} = & 2\alpha (\kket{aa}\bbra{00} - \kket{00}\bbra{00}) \\
& - 2 \alpha \sum_{m=1}^N (\kket{0m}\bbra{0m} + \kket{m0}\bbra{0m}).
\end{split}
\label{eq:L_abs_li}
\end{equation}
Inserting this expression into eq. \eqref{eq:eta_s_intermediate}, we obtain
\begin{equation}
\begin{split}
  \eta\SC = & 2 \kappa \lim_{z \rightarrow 0} \bbra{ss}\mk \Omega\TC(z)\kket{aa} + 2 \kappa \lim_{z \rightarrow 0} \bbra{ss}\mk \Omega\TC(z)\kket{00} + \\
  & \frac{\kappa}{\alpha} \lim_{z \rightarrow 0} \frac{\bbra{ss} \mk \Omega\TC(z) \mk \KK_{\rm abs}^{\rm ph}\mk \Omega\SC(z)\kket{\rho_0}}{\bbra{00}\mk \Omega\SC(z)\kket{\rho_0}}.
\end{split}
\end{equation}
We identify the first term as the efficiency for the transient case $\mk \eta\TC$, eq. \eqref{eq:efficiency_transient}, while the second term is zero. This is evident if we convert the limiting expression to the time-dependent picture, where it corresponds to the time-integral of the the sink site population for the evolution with $\mk \Omega\TC$ and initial state $\kket{00}$. The last term is a contribution due to the effect of phonon environment on the absorption. With the corresponding stationary state we can write
\begin{equation}
\etadiff= \frac{\kappa}{\alpha} \lim_{z \rightarrow 0} \frac{\bbra{ss} \mk \Omega\TC(z) \mk \KK_{\rm abs}^{\rm ph}\kket{\rho_\infty}}{\bbrakket{00}{\rho_\infty}}.
\label{eq:eta_abs_stat}
\end{equation}

Alternatively, we shall rewrite the expression for $\etadiff$, eq.~\eqref{eq:eta_abs_stat}, so that it does not explicitly dependent on the stationary state $\kket{\rho_\infty}$ and the absorption rate $\alpha$.

First, we note that $\mk \KK_{\rm abs}^{\rm ph}$ does not contribute to the absorption, i.e.
\begin{equation}
\mk \KK_{\rm abs}^{\rm ph}\kket{00} = 0.
\label{eq:kernel_no_abs}
\end{equation}
This can be seen if one writes expression for singular and non-singular part of the kernel from eqns. \eqref{eq:KKs} and \eqref{eq:KKns} for all DOFs (system and environment), i.e., full Liouvillian $\LL = \LL_{\rm ppc} + \LL^{\rm eff} \SC$ takes the role of $\mathcal{A}$, and projector $\PP$ is a projector acting on a full density matrix, resulting in $\PP R(t) = \rho(t)\otimes \rho_{\rm ph}$. Acting with the corresponding expressions for the kernel on the state $\kket{00}$ and taking into account that only $\LL_{\rm abs}$ acts on the ground state, we obtain
\begin{align}
  \label{eq:KKs_00}
  \KK_{s,\rm{S}} \kket{00} = & \PP \mathcal{\LL} \PP \kket{00} = \LL_{\rm abs} \kket{00},\\
  \label{eq:KKns_00}
  \begin{split}
  \KK_{ns,\rm{S}} (t) \kket{00} = & \PP \mathcal{\LL} \GG(t) \QQ \mathcal{\LL_{\rm abs}} \PP \kket{00} = \\ & \PP \mathcal{\LL} \GG(t) \QQ \PP \mathcal{\LL_{\rm abs}} \kket{00} = 0.
  \end{split}
\end{align}
The absorption effect of a non-singular part is zero due to $\QQ \PP = 0$, while the absorption effect of a singular part corresponds to the absorption Liouvillian $\LL_{\rm abs}$. The absence of the absorption term in $\mk \KK^{\rm ph}_{\rm abs}$ then follows from the comparison with the decomposed stationary kernel $\mk\KK\SC$ in eq. \eqref{eq:stat_kernel_decomposition}.

Next, we start by inserting identity $\sum_{ij}\kket{ij}\bbra{ij}$ into the numerator of eq. \eqref{eq:eta_abs_stat}, obtaining
\begin{align}
    \etadiff & = \kappa \lim_{z \rightarrow 0} \bbra{ss}\mk \Omega\TC(z) \mk \KK_{\rm abs}^{\rm ph} \sum_{m,n=1}^N \kket{mn}\frac{\bbrakket{mn}{\rho_\infty}}{\alpha \bbrakket{00}{\rho_\infty}}.
    \label{eq:eta_abs_decomp}
\end{align}
Note that the resulting sum does not contain the ground state term $\kket{00}$ due to \eqref{eq:kernel_no_abs}. The stationary state $\kket{\rho_\infty}$ is expressed with the stationary propagator $\mk \Omega\SC$, which is decomposed according to the eq. \eqref{eq:sumK} as $
\mk \Omega\SC(z) = \mk \Omega\SC^{\rm{ph}}(z) + \mk \Omega\SC^{\rm{ph}}(z) \LL_{\rm abs} \mk \Omega\SC(z)
$
with
$
\mk \Omega\SC^{\rm{ph}}(z)=(z-\mk \KK\TC-\mk \KK_{\rm abs}^{\rm ph})^{-1}.
$
Inserting decomposed propagator $\mk \Omega\SC(z)$ into the numerator of eq. \eqref{eq:eta_abs_decomp}, we obtain
\begin{equation}
\begin{split}
\frac{\bbrakket{mn}{\rho_\infty}}{\alpha \bbrakket{00}{\rho_\infty}} = &
\lim_{z\rightarrow 0} \frac{\bbra{mn}z \mk \Omega\SC^{\rm ph}(z)\LL_{\rm abs} \mk \Omega\SC(z)\kket{\rho_0} }{\alpha \bbra{00}z \mk \Omega\SC(z) \kket{\rho_0}} \\
= & 2 \lim_{z\rightarrow 0} \bbra{mn} \mk \Omega\SC^{\rm ph}(z) \kket{aa}.
\end{split}
\label{eq:alpha_dep}
\end{equation}
In the first line, we have taken into account that the stationary state of $\mk \Omega\SC^{\rm ph}(z)$ is a ground state, i.e. $\lim_{z \rightarrow 0} \bbra{mn} z \mk \Omega\SC^{\rm ph}(z) \kket{\rho_0} = \bbrakket{mn}{00} = 0$. In the second line, we have inserted the expression for $\LL_{\rm abs}$ from eq. \eqref{eq:L_abs_li}, taking into account vanishing limiting expression $\lim_{z \rightarrow 0} \bbra{mn} \mk \Omega\SC^{\rm ph} \kket{00} = 0$ and cancel common terms in the numerator and the denominator. Plugging eq. \eqref{eq:alpha_dep} back into \eqref{eq:eta_abs_decomp}, we obtain the efficiency in Laplace picture,
\begin{equation}
\etadiff = 2 \kappa \lim_{z \rightarrow 0}\sum_{m,n=1}^N \bbra{ss}\mk \Omega\TC(z) \mk \KK_{\rm abs}^{\rm ph} \kket{mn}\bbra{mn}\mk \Omega\SC^{\rm ph}(z) \kket{aa}.
\end{equation}
Note that while the expression for $\etadiff$ does not explicitly depend on the absorption rate $\alpha$, $\mk \KK_{\rm abs}^{\rm ph}$ could in principle depend on it. We have however verified numerically that $\mk \KK_{\rm abs}^{\rm ph}$ does not depend on $\alpha$ for the microscopic model of PPC used in this work by calculating $\mk \KK_{\rm abs}^{\rm ph}$ for $\alpha$ ranging over several orders of magnitude.

\bibliography{references}

\end{document}